\documentclass[aps,prl,twocolumn,superscriptaddress,showpacs]{revtex4}

\usepackage{graphicx}
\usepackage{dcolumn}
\usepackage{bm}
\usepackage{graphicx}
\usepackage{color}
\usepackage{amsmath, amsthm, amssymb}
\graphicspath{%
    {converted_graphics/}
    {./}
}
\begin{document}

\title{Preparation and Detection of a Mechanical Resonator Near the Ground State of Motion}

\author{T. Rocheleau}
\author{T. Ndukum}
\author{C. Macklin}

\affiliation{Department of Physics, Cornell University, Ithaca, NY 14853 USA}
\author{J.B. Hertzberg}

\affiliation{Department of Physics, University of Maryland, College Park, MD 20742 USA}
\author{A.A. Clerk}

\affiliation{Department of Physics, McGill University, Montreal, QC Canada H3A 2T8}
\author{K.C. Schwab}
\email[]{schwab@caltech.edu}
\homepage[]{www.kschwabresearch.com}
\affiliation{Applied Physics, Caltech, Pasadena, CA 91125 USA}

\date{\today}

\begin{abstract}
We have cooled the motion of a radio-frequency nanomechanical resonator by parametric coupling to a driven microwave frequency superconducting resonator.  Starting from a thermal occupation of 480 quanta, we have observed occupation factors as low as 3.8$\pm$1.2 and expect the mechanical resonator to be found with probability 0.21 in the quantum ground state of motion.  Cooling is limited by random excitation of the microwave resonator and heating of the dissipative mechanical bath.
\end{abstract}

\pacs{85.85.+j, 42.50.Wk, 84.40.Dc, 85.25.-j}

\maketitle

Cold macroscopic mechanical systems are expected to behave contrary to our usual classical understanding of reality; the most striking and nonsensical predictions are states where the mechanical system is located in two places simultaneously. Various schemes have been proposed to generate and detect such states\cite{Armour:2002b,Marshall:2003} and all require starting from mechanical states which are close to the lowest energy eigenstate, the mechanical ground state.

Naively treating the motion of a mechanical resonator quantum mechanically, one finds the elementary result that the energy should be quantized: $E_n=\hbar\omega_m(n+\frac{1}{2})$, where $n$ is an integer and $\omega_m$ is the resonant frequency. In thermal equilibrium, an average occupation factor is expected to follow the Bose-Einstein distribution: ${\bar n}_{m}^T=(e^{\hbar \omega_{m}/k_BT}-1)^{-1}$ , where  $T$, $2\pi \hbar$, and $k_B$  are the temperature, Planck's and Boltzmann's constants respectively.    Cooling a resonator into the quantum regime where ${\bar n}_{m}^T<<1$ , and measuring the very small motions has been challenging for a number of technical reasons; not only are very low temperatures necessary to freeze-out the mode, but detection with sensitivity at the quantum zero-point level is required:  $ x_{zp}=\sqrt{\hbar/(2m\omega_{m})}$, where $m$ is the resonator mass.  Furthermore, this strong position measurement must not heat the mode with measurement backaction\cite{Naik:2006}.

Many strategies have been proposed\cite{Courty:2001,Hopkins:2003,Wilson-Rai:2004,Martin:2004,blencowe:014511,Marquardt:2007,Wilson-Rae:2007,Xue:2007,Tian:2009} and applied to  realize the quantum regime with increasing success.  Experiments with nano-electromechanical structures have been able to reach ${\bar n}_{m}=25$ by passively cooling a nanomechanical resonator (NR)\cite{Naik:2006}, detected with a superconducting single electron transistor\cite{LaHaye:2004}.  Researchers experimenting with opto-mechanical systems have been able to utilize ultra-sensitive optical detection and radiation pressure to both cool and detect ${\bar n}_{m}=65$ in a toroidal resonator\cite{Park:2009}, ${\bar n}_{m}=37$ in microsphere resonator\cite{Schliesser:2009}, and ${\bar n}_{m}=35$ in an optical cavity\cite{Groblacher:2009}.

\begin{figure}[tbp] 
  \centering
  \includegraphics[bb=0 0 483 260,width=3.25in,height=1.75in,keepaspectratio]{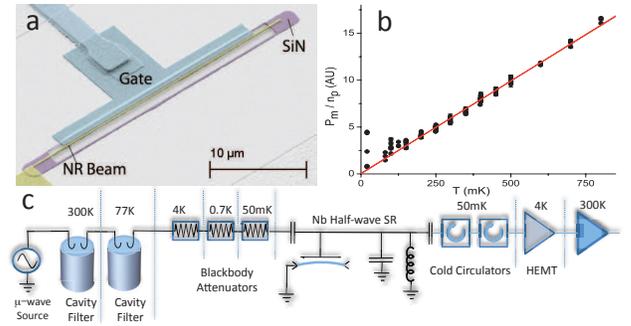}
  \caption{a) Shows the Nb/Al/SiN sample: the NR is 30 $\mu$m long, 170nm wide, 140 nm thick, formed of 60nm of stoichiometric, high-stress, LPCVD SiN\cite{Verbridge:2008} and 80nm of Al, and located 75nm from the gate electrode connected to the SR.  The SR is fabricated from a 345nm thick Nb film and has a characteristic impedance of 126$\Omega$.  (b) Shows the thermal calibration of the upconverted noise power. (c) Shows the ultra-low noise, cyrogenic measurement circuit.}
  \label{fig:fig1}
\end{figure}

The technique we employ both to cool and detect the motion of a NR close to the ground state involves parametrically coupling the motion to a superconducting microwave resonator (SR)\cite{Day:2003,Regal:2008}, (Fig. 1.)  The NR has a fundamental in-plane flexural resonance of $\omega_{m}=2\pi\cdot 6.3\text{MHz}$  and is capacitively coupled to a symmetric, two-port, half-wave SR which resonates at $\omega_{sr}=2\pi\cdot 7.5\text{GHz}$. The device is located in a dilution refrigerator and pumped through carefully filtered and cooled leads.  The thermal occupation of the SR, ${\bar n}_{sr}^T$ is expected to be 0.09 at 146mK.

The NR damping rate, $\Gamma_m^T$,  displays an unusual linear temperature dependence below 600mK, reaching $Q\sim 10^6$ at 100mK.  The SR damping rate, $\Gamma_{sr}=2\pi\cdot 600\text{kHz}$, is essentially temperature independent below 700mK and is a factor of 2.4 higher than expected from design due to internal losses.    
 
The Hamiltonian which describes the coupled resonators is given by\cite{Marquardt:2007,Wilson-Rae:2007}:
\begin{equation*}
\hat H=\hbar(\omega_{sr}+g\hat x-\lambda \hat x^2)(\hat b^{\dagger}\hat b+\frac{1}{2})+\hbar\omega_{m}(\hat a^{\dagger}\hat a+\frac{1}{2})
\end{equation*}

where $\hat a$ ($\hat a^{\dagger }$) and $\hat b$ ($\hat b^{\dagger}$) are the NR and SR annihilation (creation)  operators.   The first term shows the pondermotive-like coupling of the $\text{SR}^\prime$s frequency to the mechanical motion: $\hat x= x_{zp}(\hat a^{\dagger}+\hat a)$  and $g=\frac{\partial\omega_{sr}}{\partial x}=\frac{\omega_{sr}}{2C_t}\frac{\partial C_g}{\partial x}$ where  $C_g(x)=450\pm50aF$ is the coupling capacitance and $C_t=260fF$ is the SR total effective capacitance. The term proportional to  $\hat x^2$ results from the electrostatic frequency pulling of the mechanical resonator by the SR\cite{Cleland:1998}, where  $\lambda=\frac{\omega_{sr}}{2 C_t}\frac{\partial^2 C_g}{\partial x^2}$ , and is responsible for parametric instabilities under certain pump configurations\cite{Hertzberg:2009}.

When pumping the SR at $\omega_p=\omega_{sr}-\omega_{m}$,  harmonic motion of the NR preferentially up-converts microwave photons to frequency $\omega_{sr}$, extracting one radio-frequency NR quantum for each up-converted microwave SR photon, a process which both damps and cools the NR motion\cite{Dykman:1978,Linthorne:1990,Marquardt:2007,Wilson-Rae:2007,Xue:2007}. This cooling process is analogous to Raman scattering and the process used to cool an atomic ion to the quantum ground state of motion\cite{Diedrich:1989,Wilson-Rae:2007}. In the sideband-resolved limit, $\Gamma_{sr}<\omega_m$, the rate of this up-conversion process is given by: $\Gamma_{opt}=4 x_{zp}^2g^2\bar n_{p}/\Gamma_{sr}$, where $\bar n_{p}$ is the occupation of the SR from the pumping.  

From detailed balance, the NR occupation factor is expected to follow: 
\begin{equation*}
\bar n_m = \frac{\Gamma_m^T \bar n_m^T + \Gamma_{opt} \bar n_{sr}}{\Gamma_m^T +\Gamma_{opt}}
\end{equation*}
 where $\bar n_{sr}= \left( \Gamma_{sr}/(4\omega_m)\right)^2 + \bar n_{sr}^T\lbrack 1+2(\Gamma_{sr}/(4\omega_m))^2\rbrack$  is the effective occupation factor of the SR when $\Gamma_{opt}<\Gamma_{sr}$\cite{Dorbrindt:2008}.  The first term in the expression for $\bar n_{sr}$ is due to the quantum fluctuations of the pump field, and the second term due to the thermal occupation of the SR, $\bar n_{SR}^T$. The expressions above show that the minimum mechanical occupation possible is the effective occupation of the SR.
 
The first realization of cooling in a parametrically coupled, electro-mechanical microwave system was with a kg-scale gravitational wave transducer \cite{Blair:1995}, cooling from ${\bar n}_{m}=10^8$ to $10^5$; cooling of an NR with an SR was recently demonstrated and achieved cooling from ${\bar n}_{m}=700$ to $120$\cite{teufel:197203}. 
 
The up-converted noise power is calibrated by applying a weak pump signal, ($\Gamma_{opt}<\Gamma_m^T$), and measuring the resulting integrated sideband power, $P_m$, versus refrigerator temperature, T (Fig. 1b.)     For temperature above $\sim 150\text{mK}$, we observe the expected behavior consistent with Equipartition and use this curve to establish the relationship between measured output noise power and $\bar n_{m}$. For temperatures below 150mK we observe fluctuations in $\bar n_{m}$ apparently due to a non-thermal, intermittent, force noise at the level of $1\cdot 10^{-18} \text{N}/\sqrt{\text{Hz}}$ which is observed in other similar samples\cite{Teufel:2008,Hertzberg:2009} and similar to anomalous heating effects in other systems\cite{Stipe:2001,Deslauriers:2006}. Furthermore, the linear temperature dependence of $\Gamma_m^T$ causes the NR to decouple from the thermal environment at the lowest measured temperatures.  

The measured signal powers are consistent with our knowledge of the attenuation and gain of our measurement circuit, and estimates of the device parameters. We find $g/2\pi = 84\pm 5 \text{kHz/nm} $, which is the largest coupling strength demonstrated to date in a system of this type. From measurements of $\omega_m$ versus $\bar n_{p}$ and pump frequency, we determine $\lambda /2\pi = 2.1\pm 0.7 \text{kHz/(nm)}^2$. 

Figure 2 shows the central result of this work.  With the refrigerator stabilized at T=146mK ($\bar n_m^T=480$) we measure $\Gamma_m$ and $\bar n_{m}$ versus the SR pump occupation, $\bar n_{p}$. As is clear from Fig. 2, this process dramatically cools the motion.   However, we also observe that the SR becomes increasingly excited as $\bar n_p$ is increased(Fig. 4).  

\begin{figure}[tbp] 
  \centering
  \includegraphics[bb=13 7 761 571,width=3.25in,height=2.45in,keepaspectratio]{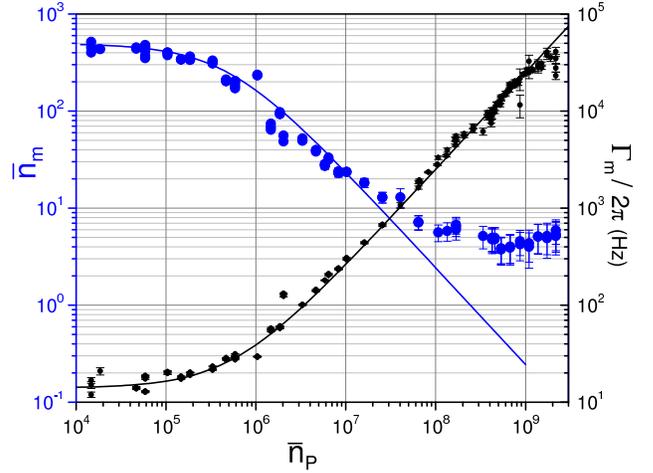}
  \caption{The figure shows $\bar n_{m}$ (\textcolor{blue}{$\bullet$}) and $\Gamma_m=\Gamma_m^T+\Gamma_{opt}$ ($\bullet$) versus $\bar n_p$.  The solid black curve is a fit to the measured $\Gamma_m$.  The solid blue curve is the expected value of $\bar n_{m}$ assuming ideal values of $\bar n_{SR}$ and $\dot n_T=3\cdot 10^4$ quanta/sec.}
  \label{fig:Fig2}
\end{figure}

Figure 3 shows the measured output noise spectra, $S_{x}(\omega)$, which is composed of up-converted microwave photons due to $\bar n_m$,  SR noise due to $\bar n_{sr}$, and HEMT amplifier noise.   Correlations between the NR motion and the SR field are important in our measured noise spectra at the lowest mechanical occupation factors.  Fluctuations in the SR voltage, due to $\bar n_{sr}$, together with the pump, produce forces at the $\omega_m$.  The resulting motion, together with the pump, produces noise at $\omega_{sr}$, however $180^o$ out of phase with the original SR fluctuations.  This correlation results in an inverted noise peak\cite{Hertzberg:2009}, similar to noise squashing\cite{Poggio:2007}, which adds incoherently to the noise power driven by the thermal bath. Our analysis shows that the  NR occupation factor is given by $\bar n_{m}=\bar n_{eff}+2\bar n_{sr}$, where $\bar n_{eff}$ is the occupation measured directly from the integrated noise peak (or dip) in the output noise spectrum, $S_{x}(\omega)$. Figure 3 shows measurements of $S_{x}(\omega)$ in three cases at low occupation factors: when $\bar n_{eff}>0$, when $\bar n_{eff}\approx 0$,  and when $\bar n_{eff}<0$ showing the squashed output noise. 

Taking the effects of $\bar n_{sr}$ into account in this way, the lowest mechanical occupation we have observed is ${\bar n}_{m}=3.8\pm 1.2$, shown in Fig. 3, with the uncertainty dominated by the uncertainty in $\bar n_{sr}$.  At this low occupation factor the resonator is expected to be found in the ground state with probability $P_0 = 1/({\bar n}_{m}+1) = 0.21$. The cooling power of this refrigeration technique is $\dot Q = \hbar\omega_m\cdot\Gamma_{opt} =10^{-22}\text{W}$

We have lowered the refrigerator temperature to 20mK and do not observe an decrease in the minimum ${\bar n}_{m}$.  Using the detailed balance relationship and the measured $\bar n_m$ and $\Gamma_{sr}$, we can compute the bath heating rate, $\dot n_T = \Gamma_m^T \bar n_{m}^T$, versus $\bar n_p$, (Fig. 4.)  It is clear that as $\bar n_p$ increases above $3\cdot 10^7$, $\dot n_T$ begins to increase, nullifying the benefit of starting at low temperatures. This level of heating is consistent with ohmic losses in the metal film on top the NR, and the thermal conductance of a normal-state electron gas.

\begin{figure}[tbp] 
  \centering
  \includegraphics[bb=99 74 668 565,width=3.25in,height=2.8in,keepaspectratio]{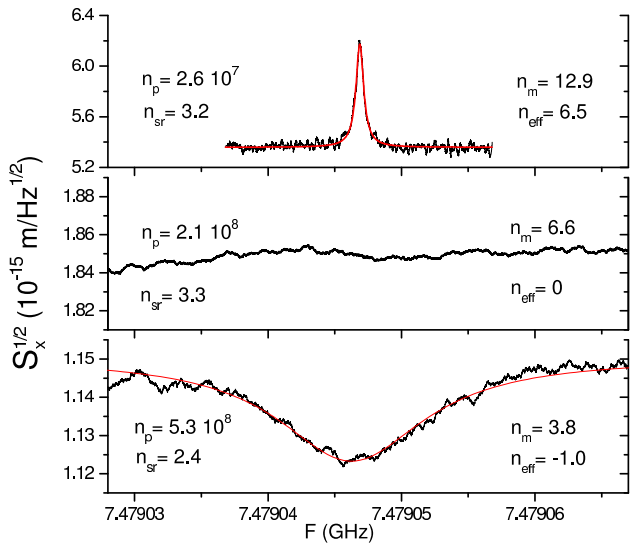}
  \caption{shows the noise squashing effect on $S_{x}(\omega)$ due to the finite occupation of the SR, in three situations: when $\bar n_{eff}>0$ (top), when $\bar n_{eff}\approx 0$ (middle), and when $\bar n_{eff}<0$ (bottom).  The red curves show Lorentzian fits through the mechanically up-converted side-band.}
  \label{fig:Fig3}
\end{figure}

\textit{Current Limitations and Future Directions:} 
These measurements identify three effects which work against the cooling process: excess fluctuations of the SR ($\bar n_{SR}$), heating of the NR thermal bath at high pump powers, and the non-thermal force noise at low temperatures.

\begin{figure}[tbp] 
  \centering
  \includegraphics[bb=68 332 663 826,width=3.25in,height=2.7in,keepaspectratio]{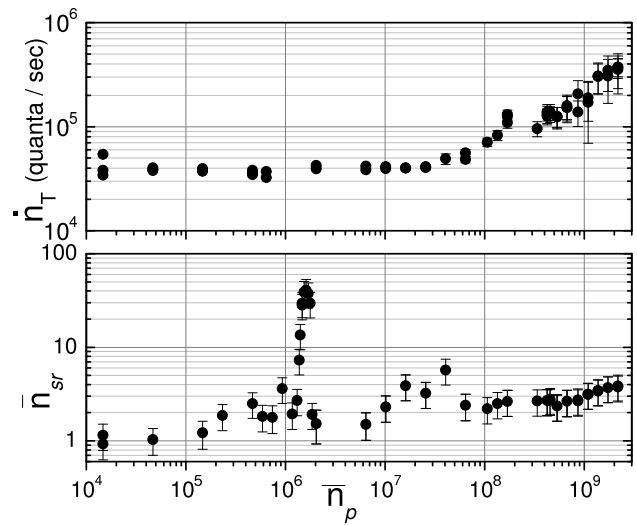}
  \caption{The upper figure shows the bath heating rate, $\dot n_T$,  versus pump strength $\bar n_p$, and the onset of excess heating above $\bar n_{p}=3\cdot 10^7$.  The lower figure shows the measured value of $\bar n_{sr}$ versus $\bar n_{p}$; the structure is suspected to be related to temporal dynamics of the transition between superconducting and normal states of the metal films and resulting microwave side-band generation\cite{Arbel-Segev:2007}.}
  \label{fig:Fig4}
\end{figure}

We believe that the excess SR occupation, $\bar n_{sr}$, is not a result of phase or amplitude noise of our microwave source: the pump signal is filtered with tunable, copper microwave cavities (one at 300K (Q=$9.5\cdot 10^3$) followed by a second at 77K (Q=$2.6\cdot 10^4$)) achieving better than $\cal L $$(+6.3MHz)< -195 \text{db}_c/\text{Hz}$, and contributing less than 0.04 photons into the SR at our highest value of $\bar n_p$.  Without these cavities the SR would be excited to $\bar n_{sr}=35$. We also believe that this excess SR occupation is not due to ohmic heating of and resulting thermal radiation from the cyrogenic attenuator network since  ${\bar n}_{sr}$ increases only weakly over a wide span of $\bar n_p$. Tests of Nb SR devices at 1.2K before the surface micromaching of the NR do not show excess dissipation and suggest that the excess losses are related to our fabrication process. 

Increasing $\Gamma_{opt}$ by engineering larger coupling strength, $g$, and/or decreasing $\Gamma_{sr}$  should be very beneficial since it will
 lead to higher cooling rates at lower pump powers, minimizing the effect of excess bath heating, $\dot n_T$.  By increasing $\Gamma_{opt}$ a factor of 10, maintaining the same $\dot n_T$, we  expect ${\bar n}_{m}\approx 0.5$, with $P_0=.67$. This approach will be limited when  $\Gamma_{opt}$ becomes comparable to $\Gamma_{sr}$ which limits the rate of cooling\cite{Grajcar:2008,Dorbrindt:2008}.   

The deep quantum limit, $\bar n_{m}\ll 1$, will be accessible when it is possible to utilize lower refrigerator temperatures and lower mechanical damping rates at these temperatures. Understanding and eliminating the excess bath heating and the non-thermal force noise will be required. Furthermore, superconducting metals on the NR also appear to be required due to the expected mechanical force noise from transport and electron momentum scattering in diffusive conductors\cite{Shytov:2002,Truitt:thesis}.  We estimate that this heating mechanism will limit $\bar n_m >3$ at $\bar n_p=3\cdot 10^8$ assuming our current device parameters and a resistance of $100\Omega$ through the NR.

These measurements show that detection with sensitivity to resolve motions approaching the ground state is possible with existing HEMT-based amplifiers.  Eliminating internal SR losses, unbalancing the SR couplings, and implementing improved microwave amplifiers\cite{Muck:2003,Castellanos:2008} would significantly reduce the measurement time.

Nonetheless, the production and detection of a NR with $\bar n_m=3.8$ is sufficient to enable future experiments. Due to Uncertainy Principle fluctuations of the mechanical motion and resulting spontaneous emission, the rate of microwave photon up-conversion is expected to differ from the rate of down-conversion.  This difference can be used as a fundamental thermometry technique\cite{Diedrich:1989,Marquardt:2007,Wilson-Rae:2007}, and would be the first quantitative measurement of the zero-point motion of a mechanical structure.
 
This level of cooling is essential to realize entangled states between superconducting quantum bits and the motion of a nanomechanical device\cite{Armour:2002b,Utami:2008,LaHaye:2009}.  Similar to procedures in atomic physics,  such an experiment would involve preparing the cold state of the mechanical device and after the refrigeration is complete, the cooling can be turned off.  The state of the cold beam could then be manipulated before thermallization of the motion.  In our realization, we expect cooling from $\bar n_m=500$ to 4 quanta in $\sim 200\mu \text{s}$, and one thermal quantum to enter the resonator in $\tau=(\dot n_{T})^{-1}=2\mu \text{s}$ which exceeds superconducting qubit manipulation times and is comparable to qubit measurement and relaxation times.

\begin{acknowledgments}
We acknowledge helpful conversations with M. Aspelmeyer, R. Ilic, M. Skvarla, M. Metzler, M. Shaw and assistance from M. Savva, S. Rosenthal, and M. Corbett. This work has been supported by the Fundamental Questions Institute fqxi.org (RFP2-08-27), and the US NSF (DMR-0804567).  Device fabrication was performed at the Cornell Nanoscale Facility, a member of the NNIN (NSF Grant ECS-0335765).
\end{acknowledgments}

\bibliography{bibfile}

\end{document}